# Electrical and magnetic transport properties of $Fe_3O_4$ thin films on GaAs (100) substrate.

Ram Prakash[1], R. J. Choudhary[1], L. S. Sharath Chandra[1],

N. Lakshmi[2] and D. M. Phase[1a]

[1]UGC-DAE Consortium for Scientific Research, University Campus Khandwa road, Indore, (M.P.) –452017 (India).

[2]Department of Physics, M.L. Sukhadia University, Udiapur, 313001 (India).

## Abstract

Thin films of magnetite ($Fe_3O_4$) are grown on single crystal GaAs (100) substrate by pulsed laser deposition. X ray diffraction (XRD) result shows the (111) preferred orientation of the $Fe_3O_4$ film and x-ray photoelectron spectroscopy confirm the presence of single phase $Fe_3O_4$ in the film. The electrical transport property of the film shows the characteristic Verwey transition at 122 K and below 110 K, the transport follows variable range hopping type conduction mechanism. The film shows room temperature magnetization hysteresis loop suggesting the ferrimagnetic behavior of the film with saturation magnetization value close to 470 emu/cc.

---

[a] Corresponding author e mail: dmphase@csr.ernet.in

**Introduction**

Ever since the magnetism in magnetite ($Fe_3O_4$) was discovered, the material revolutionized the world of science with its fascinating properties. The material has been at the core of the tremendous applications such as electric motors, electromagnets, transformers, video/audiotapes, magnetic inks and biomedical applications etc. In the recent years, there have been developments in the storage devices such as hard disc, floppy disc, read heads, bubble memory, magnetic random access memory (MRAM) etc. The development of this field is attributed to the search of spintronic-based devices [1-3], wherein spin based transport properties of a material are exploited. $Fe_3O_4$ also falls under the category of such a functional material owing to its several fascinating properties such as high Curie temperature (~850 K), half metallicity, and low electrical resistivity at room temperature.

$Fe_3O_4$ is an inverse cubic spinel compound, wherein iron ions are shared between tetrahedral and octahedral sites; tetrahedral sites being occupied by $Fe^{3+}$ ions whereas, octahedral sites by both $Fe^{3+}$ and $Fe^{2+}$ ions in the same ratio. The magnetic interaction among iron ions at octahedral and tetrahedral sites is antiferromagnetic and that among octahedral ions is ferromagnetic; overall a ferrimagnetic arrangement of $Fe_3O_4$. Therefore, the net magnetic moment in $Fe_3O_4$ is due to $Fe^{2+}$ ions (4 $\mu_B$). $Fe_3O_4$ undergoes charge ordering at 120 K (known as Verwey transition temperature, $T_V$) leading to an abrupt increase in its resistivity while cooling [4]. Across this temperature, structural transition also takes place from high temperature cubic structure to a low temperature monoclinic structure. In spite of evident spintronic potential of $Fe_3O_4$, it has been so far not been optimized for its best functionality in the field. The crucial reason for this lack of its inclusion in device architectures is rather poor control and exploration of the growth of $Fe_3O_4$ films on the technologically important semiconducting substrates like Si or GaAs. In this context, we attempt to study the nature of $Fe_3O_4$ thin film grown on GaAs substrate and compare it with the films grown on conventionally used MgO substrate for $Fe_3O_4$.

Thin films of magnetite have been prepared by variety of deposition techniques such as molecular beam epitaxy (MBE), reactive magnetron sputtering, pulsed laser deposition etc on different substrates like MgO, $MgAl_2O_4$, $\alpha$-$Al_2O_3$, $SrTiO_3$, Pt, Si, GaAs etc [5-15] for their transport and magnetic properties. From these studies there is a consensus that the electrical transport properties vis a vis the magnetic properties of magnetite thin films are function of the nature of defects (internal or external) and defects density present in the film. From the available literature, we note that the growth of $Fe_3O_4$ films on GaAs substrate is not well explored, though it has huge potential in spintronic devices. Lu et. al. [[12]] employed electron beam evaporation to grow $Fe_3O_4$ films by initially growing Fe films on GaAs (100) followed by its oxidation. The procedure resulted in the epitaxial growth of magnetite on GaAs (100) with epitaxial relation as $Fe_3O_4$ (100)<011>// GaAs(100)<010>. However, such a method limited the thickness of the film to 6 nm only. They also showed that easy axis of magnetization is along [0 -1 -1] direction of GaAs (100) substrate. Watts et al.[13,[14]] reported the (111) preferred oriented $Fe_3O_4$ thin film on GaAs (100) substrate but they detected the presence of some other orientations as well, such as (220), (311), (511) etc. Kennedy et al. [[15]] reported (111) oriented magnetite film using iron target for ablation in presence of $O_2$ background by Nd:YAG pulsed laser deposition technique. Unusually high magnetization (730 emu/cc) was observed in the film because of amorphous iron present in the film. From these studies it seems that a general agreement regarding the properties and the growth of $Fe_3O_4$ films on GaAs substrate is yet to be achieved. In the present paper we study the structural, electrical, and magnetic transport properties of $Fe_3O_4$ thin film grown by pulsed laser deposition technique on GaAs (100) substrate.

**Experimental Details**

The thin films of $Fe_3O_4$ were grown on single crystal GaAs (100) by pulsed laser deposition (PLD) technique. The target used for the deposition is $\alpha$-$Fe_2O_3$ pallet prepared from high purity $\alpha$-$Fe_2O_3$ powder, sintered at 950 °C for 24 hours. Before ablation surface contaminations were removed from the GaAs substrates using a $H_2SO_4$: $H_2O_2$: $H_2O$ (4:1:1) solution etching for 30 seconds, subsequently rinsed by de-ionized water and methanol cleaning. The cleaned substrate was mounted on a heater (substrate holder) and placed immediately in the deposition chamber, which was then evacuated to the base pressure of $2 \times 10^{-6}$ Torr. During the ablation, the laser beam from a KrF excimer laser (wavelength 248 nm) was focused on to the target and its repetition rate was kept at 10 Hz. The laser energy density at the target was kept at 1.8 J/cm$^2$ and the substrate temperature was maintained at 450° C during deposition. The deposition was carried out for 20 min. The thickness of the film was 1500 Å, as measured by stylus profilometer (Ambios, USA). These films were characterized by x-ray diffraction (Rigaku, Japan), and magnetization measurements. The electronic state of the iron ions and the presence of any other phase in the film were examined by core level x-ray photoelectron spectroscopy (Omicron-EA 125) using Al-K$_\alpha$ (1486.6eV) radiation source. The spectral resolution of spectrometer is 0.8 eV. Electrical and magnetic transport measurements were performed using physical property measurement system (PPMS-Quantum Design, USA) by standard four probe method. The room-temperature magnetic hysteresis behavior of these thin films was examined using the vibrational sample magnetometer (VSM) technique (Lakeshore, Model 7401).

**Results and discussions**

The structure of the film was examined by XRD using Cu-K$_\alpha$ radiation in $\theta$ -2$\theta$ geometry. We observed that films are highly oriented along (111) direction with cubic structure (Fig. 1). The lattice parameter of the film, as calculated from the XRD pattern, is found to be 8.370 Å, which is close to the bulk lattice parameter (8.393 Å). It is recalled here that Lu et al. [12] observed the epitaxial growth of Fe$_3$O$_4$ on GaAs (100) substrate with epitaxial relation as Fe$_3$O$_4$ (100)<011>// GaAs(100)<010>. Other reports such as Watts et al. [[14]] and Kennedy et al. [[15]] suggested the (111) preferred oriented growth of the film as observed in the present study. However, Watts et al. detected the minor contribution of other planes in their films. Whereas, Kennedy et al. [15] suggested the possibility of some amorphous unoxidized iron at the interface of the film and substrate, since they used iron target for ablation in presence of O$_2$ background for reactive deposition of Fe$_3$O$_4$. One possible reason for the (111) orientation of the Fe$_3$O$_4$ film on GaAs (100) substrate may be that (111) oriented Fe$_3$O$_4$ surface is the most energetically favorable one since the <111> direction has the highest areal atomic density in the Fe$_3$O$_4$ spinel crystal structure [[16]]. It is noted here that in the present case the lattice mismatch between the film and substrate materials is very large, therefore, the growth orientation will be determined by the thermodynamically stable state having minimum internal energy. This is further confirmed by our very recent study [[11]] of pulsed laser deposited $Fe_3O_4$ films on Si substrate, where we have shown that the films are (111) orientated, independent of the Si substrate orientations.

Since the lattice parameters of $Fe_3O_4$ (0.8396 nm) and $\gamma$-$Fe_2O_3$ (0.8342 nm) are very close to each other, XRD pattern cannot distinguish between these two phases of iron oxide. To confirm the growth of $Fe_3O_4$ film, the sample was characterized by X-ray photoelectron spectroscopy (XPS). We sputtered the film to remove the surface contamination and to understand the chemical state of iron deep inside the film. Sputtering was performed at very low energy (500eV) of $Ar^+$ ion so that it does not influence the composition of the film. In Fig. 2 (a) we show the survey scans of as deposited and 30 min sputtered films. It is evident that as-deposited film has small surface contamination of carbon, which is removed by sputtering. The XPS spectrum of sputtered film has peaks corresponding to iron and oxygen. For the analysis of chemical compositions, the narrow scans were recorded for Fe and O core levels. In Fig. 2 (b) we show the Fe 2p core level spectra of the as deposited and 30 min sputtered film recorded in the binding energy (B.E.) range of 731 to 703 eV. These spectra were normalized with maximum intensity and shifted up side vertically for sake of clarity. Because of the surface contamination in the as grown film, we observe a lesser intense spectrum but after etching the film with $Ar^+$ ion, the intensity of the Fe 2p core level improves. From the Fig. 2 (b), we observe that the Fe $2p_{1/2}$ and $2p_{3/2}$ peaks situated at around 711 and 724 eV are broadened due to the existence of the both $Fe^{2+}$ and $Fe^{3+}$ ions. The Fe $2p_{3/2}$ and Fe $2p_{1/2}$ binding energies for $Fe^{2+}$ and $Fe^{3+}$ were determined by fitting the spectral line shapes to a convolution of Gaussian and Lorentzian functions (inset of Fig. 2b). The representative spectrum of 30 min sputtered sample is shown in the inset of Fig. 2(b). The measured Fe $2p_{3/2}$ (Fe $2p_{1/2}$) binding energy is 709.2 (722.4eV) for $Fe^{2+}$ and 711.1 eV (724.5eV) for $Fe^{3+}$. These values match very well to the literature values.[18] From these fitted parameters we have calculated $Fe^{+3}/Fe^{+2}$ ratio, which comes out to be 2:1 as expected for $Fe_3O_4$. Since the binding energy position of satellite of Fe $2p_{3/2}$ for different Fe oxidation states +2 or +3 occurs at 715 or 719 eV respectively, the corresponding satellite structure can be used to provide information about the presence of FeO or γ-$Fe_2O_3$. $Fe_3O_4$ being the mixed state of FeO and $Fe_2O_3$, its spectra has smeared out unresolved structure in this energy range [[19]-[21]]. The features of the core level confirm the presence of $Fe_3O_4$ phase in the film and absence of any other phase of iron oxide. Similar spectra are obtained for the films etched for longer duration confirming the $Fe_3O_4$ phase deep inside the film as well.

After ensuring the structure and single phase nature of $Fe_3O_4$ film, the resistivity measurement (ρ-T plot) was performed using standard four probe method down to 66 K {Fig. 3(a)}. We notice that at 122 K, resistance increases sharply as we lower the temperature, a signature of the disorder to order transition. However, the transition is not that sharp as generally observed for bulk samples. The room temperature resistivity value of the film (10 mΩ cm) is higher than the single crystal bulk sample (4 mΩ cm). The higher value of resistivity and a rather broader transition attributed to the incoherent growth of the films resulting from huge lattice mismatch between the films and the substrates. The activation energy of the film is calculated using Arrhenius equation ($\rho_H = \rho_0 \exp(-E_g/kT)$) beyond Verwey transition temperature in the temperature range of (130 – 300 K). Fig. 3(b) shows the plot between ln$\rho$ and 1/T. The plot shows a linear behavior in the studied temperature range and the activation energy is calculated to be 62.2 meV, which is close to the reported data for $Fe_3O_4$ bulk value (58 meV) [[22]].

To understand the transport behavior below $T_V$, we fit our resistivity data with Mott variable range hopping (VRH) model wherein conduction is considered by hopping between localized states, represented by expression [[23]] $\rho = \rho_0 \exp(T_0/T)^{1/4}$, where $\rho_0$ depends on phonon density and $T_0$ is characteristic Mott temperature, a parameter determining degree of disorder given by $k_B T_0 = 18\alpha^3/N(E)$, $\alpha$ being the inverse of localization length and N(E) being density of states at Fermi level. In the upper inset of Fig. 3(b), we show the plot between ln $\rho$ vs $(1/T)^{0.25}$ in the temperature range of 115 K to the lowest possible measured temperature 66 K. (Below 66 K, film becomes too insulating to measure). It is evident from the plot that ln $\rho$ fits linearly with $(1/T)^{0.25}$ in the studied temperature range below $T_V$. From the slope of the linear fitting, we calculate the value of $T_0$ as $1.1 \times 10^8$ K, which is close to the earlier reported value ($1.6 \times 10^8$ K) for epitaxial films of $Fe_3O_4$ on different substrates [[10], [24]]. However, the value of localization length (1/$\alpha$) is calculated to be 0.064 nm. This value of localization length is not physically meaningful. Therefore, considering the fact that below $T_V$, $Fe_3O_4$ has charge ordered state, we used Shklovskii and Efros VRH model (SE-VRH), [[25]] which accounts for electron electron Coulombic interaction term. It is recalled here that there is a close similarity between the transport properties of charge ordered manganites and magnetite, since in the both the cases Coulomb interaction plays and important part on the hopping conduction of electrons. Hence, SE model developed for the temperature dependence of resistivity in manganites by considering the theory of weak localization and VRH in the presence of a Coulomb gap, should be applicable for $Fe_3O_4$ as well. SE-VRH model is expressed by $\rho = \rho_0 \exp(T_0/T)^{1/2}$, where $T_0 = 2.8e^2\alpha/(4\pi k\varepsilon_0)$, k being the dielectric constant of the material. Figure 4(b) shows the plot between $\ln \rho$ vs $(1/T)^{0.5}$ in the temperature range of 110 K to the lowest possible measured temperature 66 K. The slope of the linear fitting in the plot yields $T_0 = 23440$ K. Using this value of $T_0$, we estimate the coherence length ($1/\alpha$) equals to ~ 2 nm, which is almost seven times of the distance between nearest neighborhood hopping distance between $Fe^{+2}$ and $Fe^{+3}$ ions at octahedral B sites of $Fe_3O_4$. This value of coherence length as calculated from the SE-VRH model is physically reasonable. Therefore, it seems that the SE-VRH model fits for the $Fe_3O_4$ film below $\sim 0.9 T_V$ down to measureable temperature of $\sim 0.5\ T_V$.

Now if we consider that below $T_V$ the transport behavior shows SE-VRH type behavior then, the SE-VRH energy ($\Delta$) should be equal to the Coulomb soft gap (U) occurring because of strong electronic correlation. To validate this point, we calculate the $\Delta$ using formula $\Delta = k_B\sqrt{(T_C T_0)}$, as derived from SE-VRH model. Here $T_C$ is the materials characteristic temperature where the transition from nearest neighbor thermally activated hopping transforms into SE-VRH type hopping. In our study we consider $T_C$ as equal to 110 K, which is the maximum temperature for SE-VRH behavior. Using the calculated values of $T_0$ and $T_C$ from the graph, we estimate $\Delta = 0.138$ eV. U can be calculated using standard electrostatic relation $U = e^2/(4\pi\varepsilon_0 R_{avg})$, where $R_{avg}$ is the average distance between hopping sites [26]. Taking $R_{avg}$ value to be 10 nm as calculated from the Hall measurement [[27]] and putting the respective values of $T_0$, e and $\varepsilon_0$, we obtain U as equal to 0.144 eV. The observed values of U and $\Delta$ are very close to each other suggesting the applicability of SE-VRH model below $T_V$ in the present study. Earlier reports suggested the Mott VRH type conduction behavior below $T_V$ either in single crystal or epitaxial films. We believe that in our study though Mott VRH model fits well, SE VRH behavior is more justified, given the fact that our films are not epitaxial but well oriented. Since in the present study the grain density would be larger than the epitaxial films, the transport behavior should occur via tunneling across the interface of two adjacent grains, which would provide a linear behavior between ln $\rho$, and $(1/T)^{0.5}$ [[28]].

Figure 4 shows the magneto resistance (MR%) behavior with magnetic field at different temperatures (90K, 150K and 300K), MR% is defined as $(R_H-R_0)*100/R_0$. The magnetic field was applied perpendicular to the film plane. It may be seen that MR% is higher below $T_v$ than room temperature value. We fit our MR data with expression $\rho_H = \rho_0 [1+\alpha H+\beta H^2]$, which has been earlier used for (111) oriented epitaxial thin film of $Fe_3O_4$ deposited on Sapphire substrate [[10]]. The MR data at all three temperatures fit very well to this form. At 90 K temperature the values of $\alpha$ and B are 0.0108 $T^{-1}$ and $3.1\times10^{-3}$ $T^{-2}$ respectively. Ogale et al explained the similarity of field dependent behavior of magnetoresistance between the magnetite and manganites in terms of the importance of small polaron in the conductance behavior of these two systems.

In Fig. 5 we show the room temperature magnetization hysteresis behavior of the film. It is evident from the figure that the film shows a clear magnetic hysteresis behavior with coercivity of 372 Oe at room temperature. The magnetization saturates at field value of 0.3 T with saturation magnetization (Ms) value close to 470 (±20) emu/cc, which is slightly lower than the single crystal saturation value (480 emu/cc) [[5]]. The observed behavior suggests the room temperature ferrimagnetic behavior of the film. In the earlier report, Kennedy et al. observed higher $M_S$ value of 730 emu/cc in their $Fe_3O_4$ film on GaAs substrate, which they accounted to the presence of metallic Fe rich regions within the film. The minor deviation of $M_S$ value from that of the single crystal may be because of the presence of growth related defects in the film. It is known that epitaxial $Fe_3O_4$ films grown on single crystal MgO substrate possess a high density of anti phase boundary (APB) defects arising because of the double lattice parameter of $Fe_3O_4$ (8.3967 Å) as compared to that of MgO substrate (4.213 Å) and lower symmetry of $Fe_3O_4$ compared to MgO. The consequence of these APBs defects brings in the abnormal electrical and magnetic properties of $Fe_3O_4$, in particular the non-saturation of magnetization even at high field. The presence of APBs and its related effects are also known to occur in polycrystalline films of $Fe_3O_4$ grown on silicon substrates. In the present study the magnetization result ostensibly implies that our film is free from APBs defects or has lesser density of such defects. However, we need a more plausible experimental verification in this regard.

**Conclusions**

In conclusion we have successfully grown $Fe_3O_4$ thin film on single crystal GaAs (100) substrate using pulsed laser deposition technique from $\alpha$- $Fe_2O_3$ target. These films are (111) oriented and single phase as confirmed by X-ray diffraction and X-ray photoelectron spectroscopy respectively. These films also show a characteristic Verwey transition at 122 K. The electrical transport behavior of the films below $T_v$ shows SE-VRH type behavior while above $T_V$ it is thermally activated type. The magnetization data reveals the room temperature ferrimagnetic behavior of the film with saturation magnetization close to the bulk value.

**ACKNOWLEDGEMENTS:**

The authors are thankful to Dr. P. Chaddah and Prof. A. Gupta for the encouragement and acknowledge Dr. V. Ganesan for providing access to magneto-resistance measurements. We also thank Shailja Tiwari for her help in the sample preparation. One of us (RP) would like to acknowledge CSIR, N. Delhi for the financial support.

**Figure Captions:**

**Figure 1:** XRD pattern of $Fe_3O_4$ thin film on GaAs (100) substrate grown by pulsed laser deposition. The film peaks are indicated by F and the reflection planes are indicated in parenthesis.

**Figure 2**: (a) Survey scan and (b) Fe-2p Core level x-ray photoelectron spectra of as-deposited and 0.5 keV $Ar^+$ ion sputtered (30 min) $Fe_3O_4$ thin film on GaAs (100) substrate grown by pulsed laser deposition. The inset in Fig. (b) shows the representative fitted spectra of Fe2p core level.

**Figure 3:** (a) Resistivity ($\rho$) as a function of temperature (T) for the $Fe_3O_4$ film; (b) Plot between $\ln\rho$ vs 1/T with a linear fitting of the plot in the temperature range of 130K-300 K; upper and lower insets show the plot between $\ln\rho$ vs $1/T^{0.25}$ and $\ln\rho$ vs $1/T^{0.5}$ with linear fitting in the temperature range of 115 K-66K and 110 K-66 K respectively.

**Figure 4.** Field dependence of Magneto resistance (MR %) of $Fe_3O_4$ thin film on GaAs (100) substrate grown by pulsed laser deposition at temperatures 90K, 150K and 300K.

**Figure 5**: M versus H plot at 300 K of $Fe_3O_4$ film on GaAs (100) substrate.

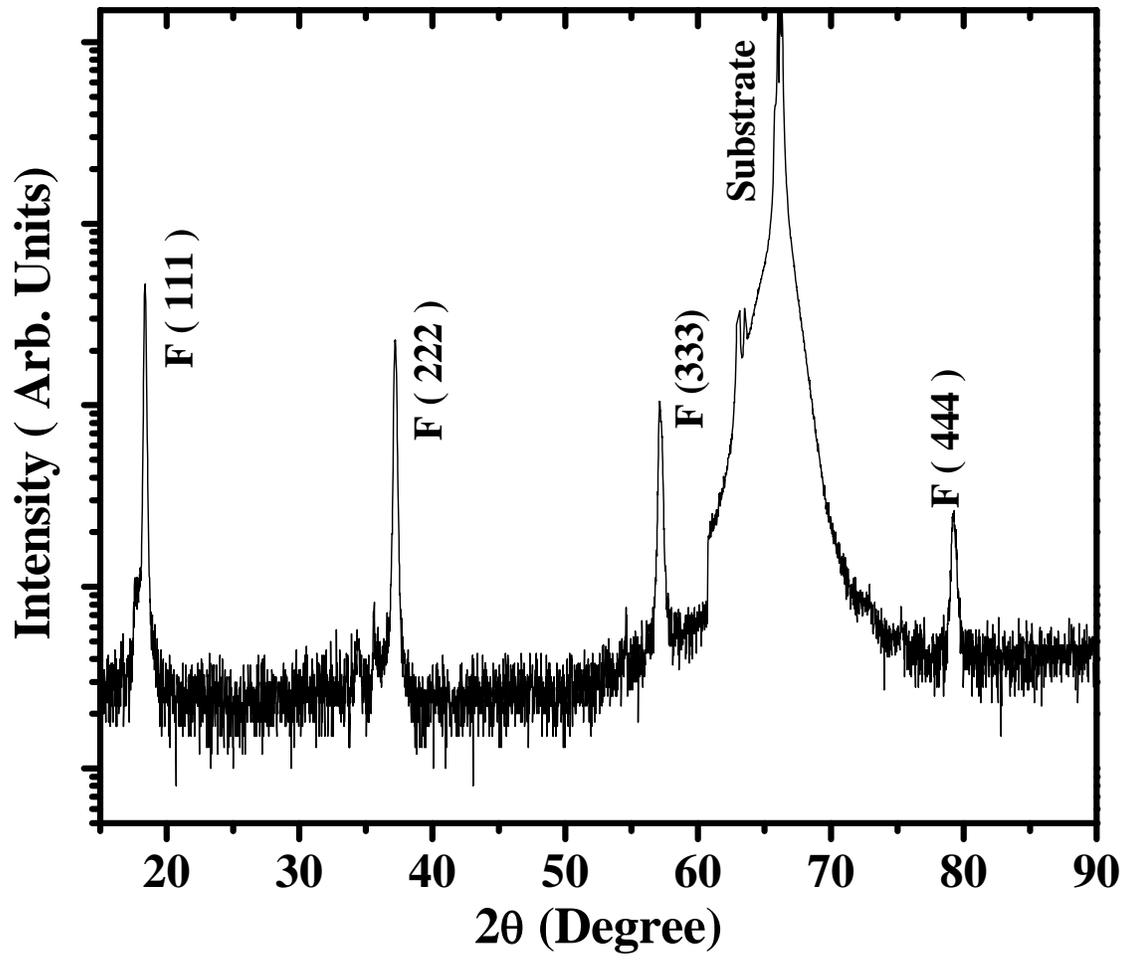

Figure 1

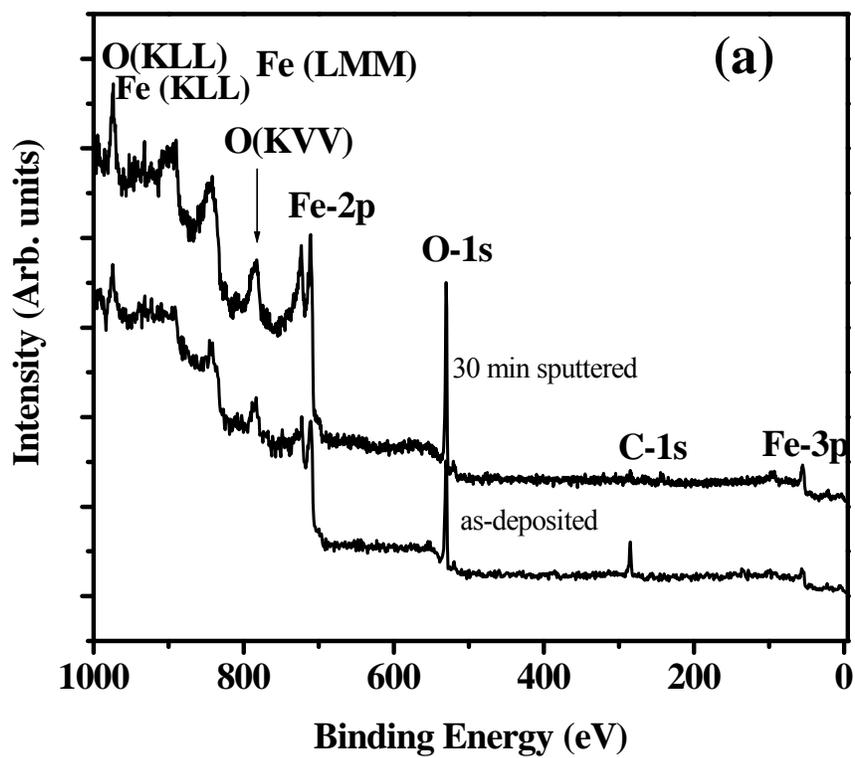

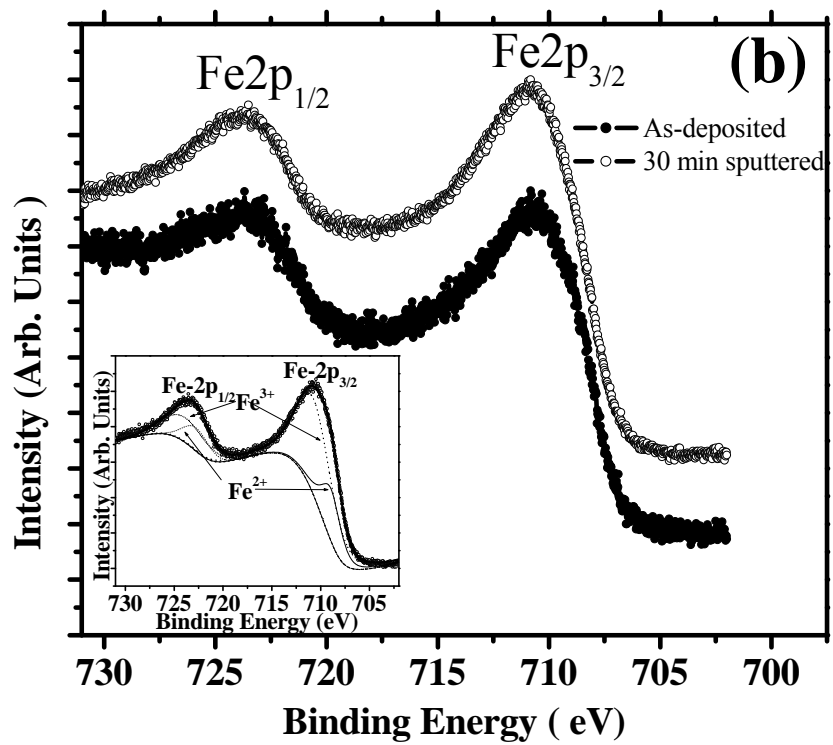

Figure 2

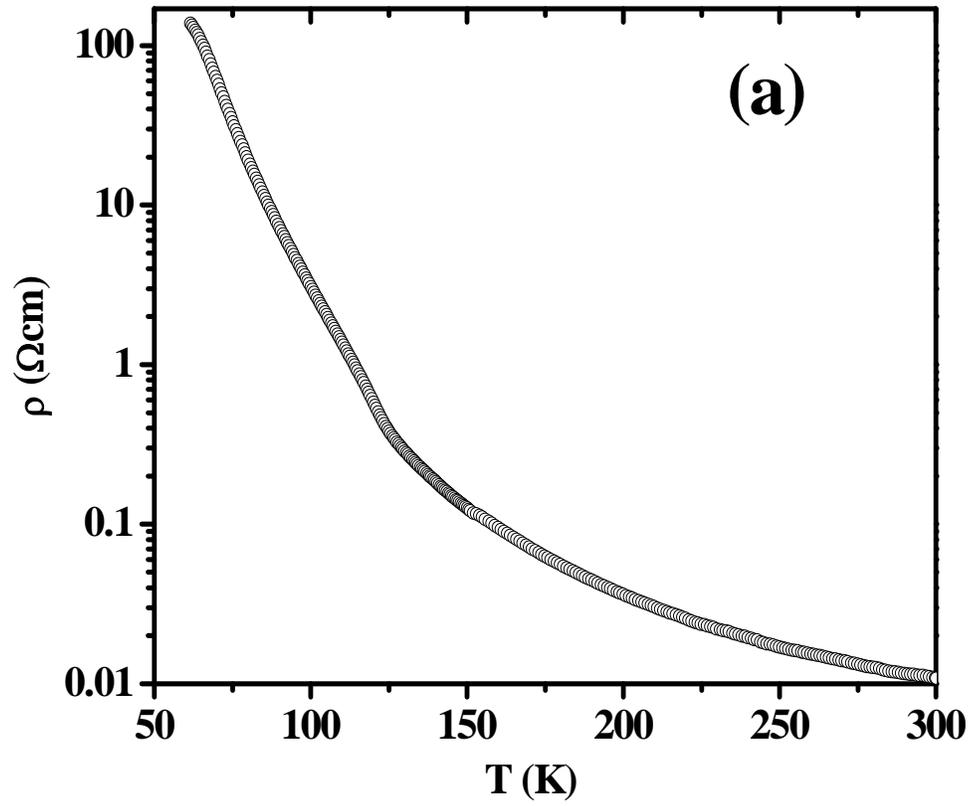

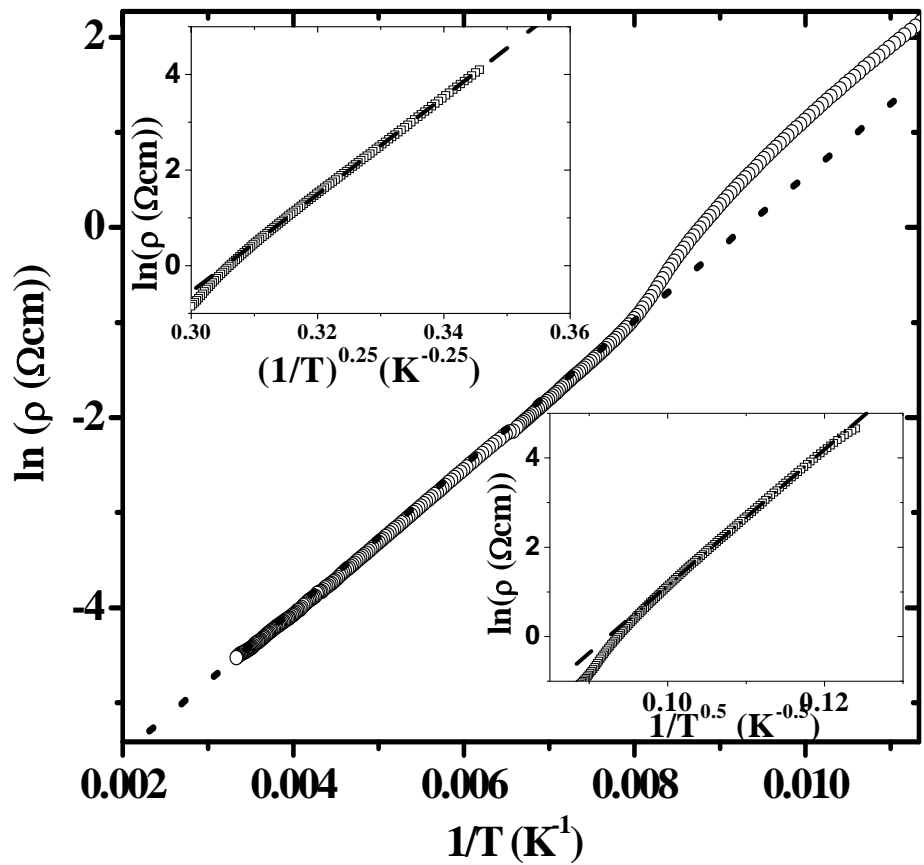

Figure 3

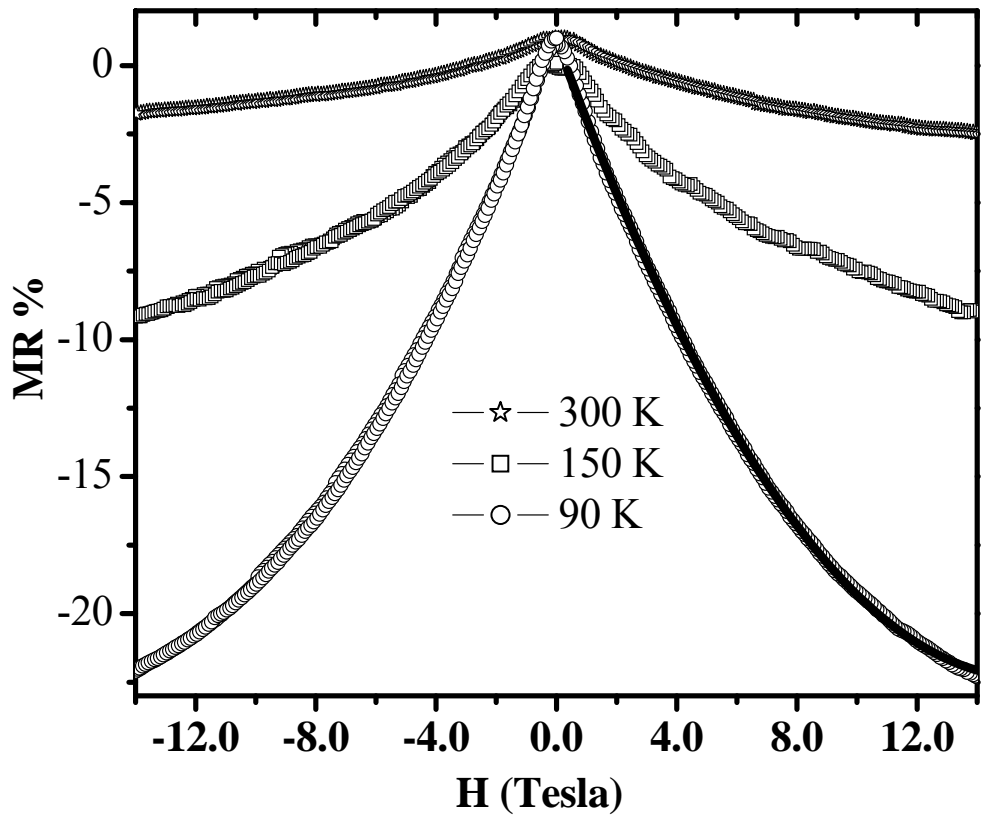

Figure 4

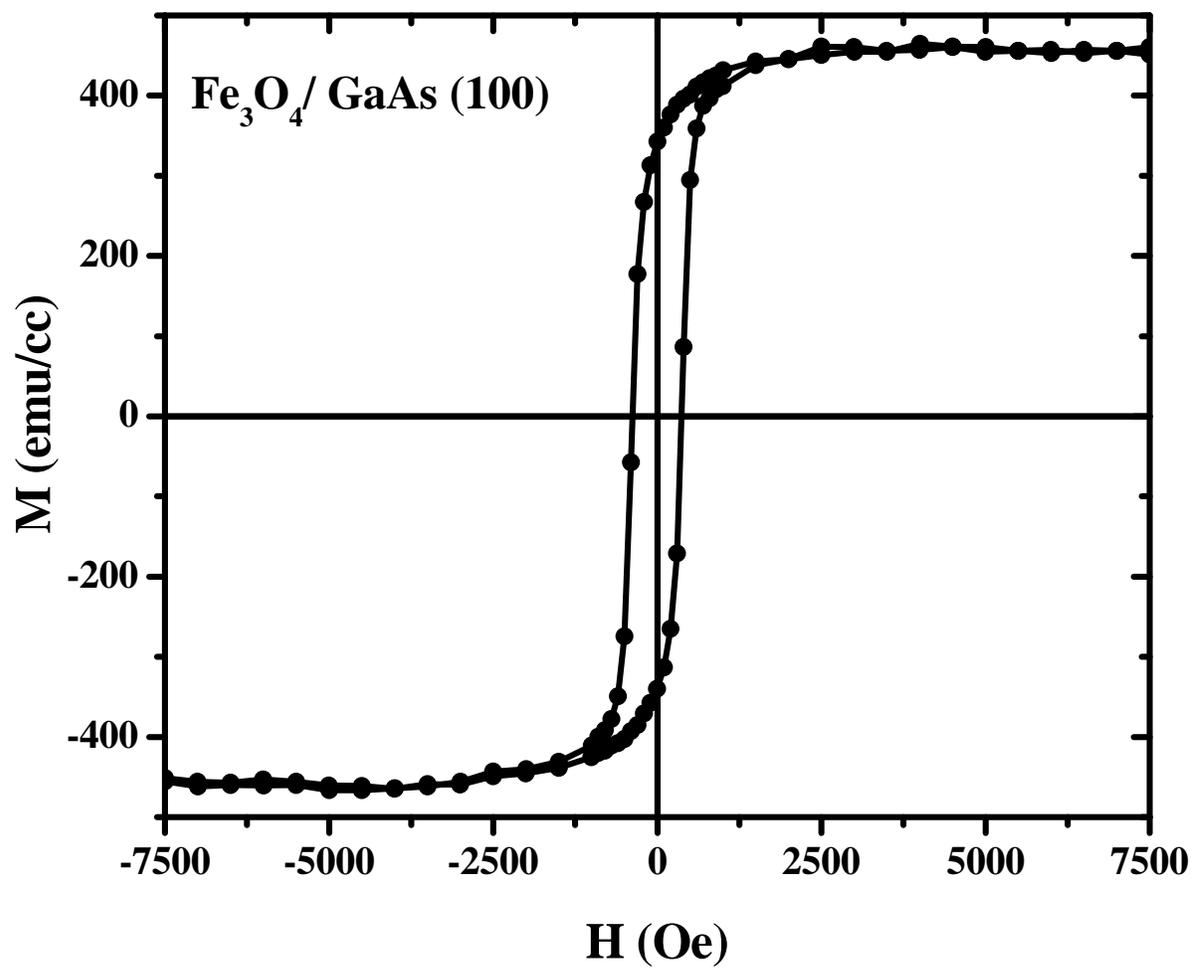

Figure 5